\documentclass[a4paper,11pt]{article}
\pdfoutput=1 % if your are submitting a pdflatex (i.e. if you have
\usepackage{jcappub} % for details on the use of the package, please
\usepackage[T1]{fontenc} % if needed
\usepackage[lofdepth,lotdepth]{subfig}

\title{Introducing the MeVCube concept: a CubeSat for MeV observations}

\author[a,1]{Giulio Lucchetta,\note{Corresponding author.}}
\author[a]{Markus Ackermann,}
\author[a,b]{David Berge,}
\author[a]{and Rolf B{\"u}hler}
\affiliation[a]{Deutsches Elektronen-Synchrotron DESY,\\Platanenallee 6, 15738 Zeuthen, Germany}
\affiliation[b]{Institut f{\"u}r Physik, Humboldt-Universit{\"a}t zu Berlin,\\Newtonstra{\ss}e 15, 12489 Berlin, Germany}
\emailAdd{giulio.lucchetta@desy.de}
\emailAdd{markus.ackermann@desy.de}
\emailAdd{david.berge@desy.de}
\emailAdd{rolf.buehler@desy.de}

\abstract{Despite the impressive progress achieved both by X-ray and gamma-ray observatories in the last few decades, the energy range between $\sim200\;\mathrm{keV}$ and $\sim50\;\mathrm{MeV}$ remains poorly explored. \emph{COMPTEL}, on-board the \emph{Compton Gamma-Ray Observatory} (\emph{CGRO}, $1991$-$2000$), opened the MeV gamma-ray band as a new window to astronomy, performing the first all-sky survey in the energy range from $0.75$ to $30\;\mathrm{MeV}$. More than $20$ years after the de-orbit of \emph{CGRO}, no successor mission is yet operating. Over the past years many concepts have been proposed, for new observatories exploring different configurations and imaging techniques; a selection of the most recent ones includes \emph{AMEGO}, \emph{ETCC}, \emph{GECCO} and \emph{COSI}.\\
We propose here a novel concept for a Compton telescope based on the CubeSat standard, named \emph{MeVCube}, with the advantages of small cost and relatively short development time. The scientific payload is based on two layers of pixelated Cadmium-Zinc-Telluride (CdZnTe) detectors, coupled with low-power read-out electronics (ASIC, VATA450.3). The performance of the read-out electronics and CdZnTe custom designed detectors have been measured extensively at DESY \cite{Lucchetta_CZT}. The performance of the telescope is accessed through simulations: despite a small effective area limited to a few $\mathrm{cm}^{2}$, \emph{MeVCube} can reach an angular resolution of $1.5^{\circ}$ and a sensitivity comparable to the one achieved by the last generation of large-scale satellites like \emph{COMPTEL} and \emph{INTEGRAL}. Combined with a large field-of-view and a moderate cost, \emph{MeVCube} can be a powerful instrument for transient observations and searches of electromagnetic counterparts of gravitational wave events.}

\begin{document}
\maketitle
\flushbottom

\section{Introduction: the ``MeV gap''}
\begin{figure}[htbp]
\centering
\includegraphics[width=0.95\textwidth]{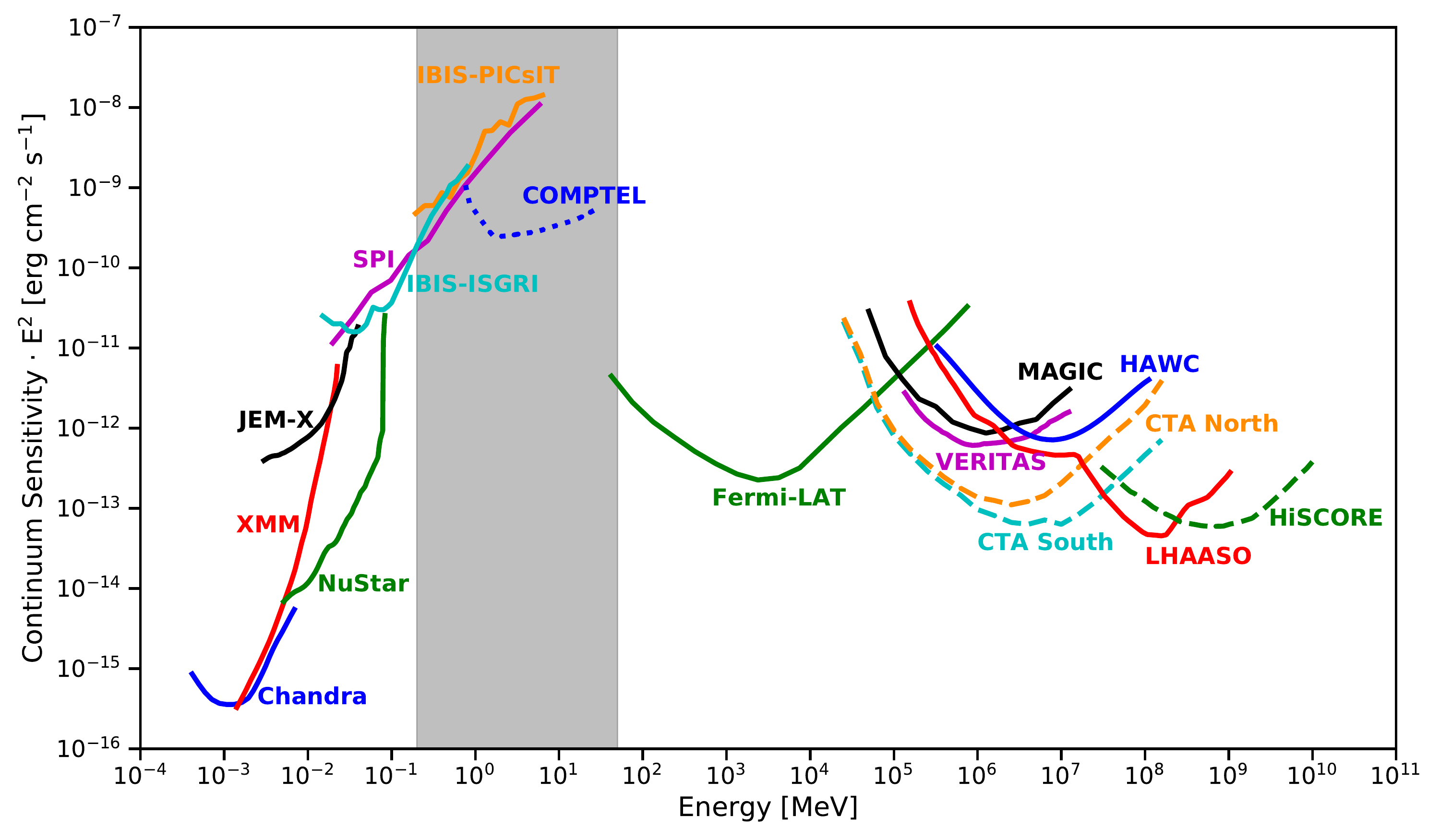}
\caption{\label{fig:MeVGap} Continuum sensitivity of different X-ray and gamma-ray instruments: past missions are shown in dotted lines, currently operating instruments in solid lines and future planned missions in dashed lines. \emph{Chandra} and \emph{XMM-Newton} sensitivities are computed for an observation time of $10^5 \;\mathrm{s}$ \cite{SensForXRays}. \emph{NuSTAR} data are taken from \cite{NuStar_Sensitivity} for an observation time of $10^6 \;\mathrm{s}$. For the \emph{INTEGRAL} observatory, the sensitivity is estimated for an observation time of $10^6 \;\mathrm{s}$ for \textit{JEM-X}\textsuperscript{1} and \emph{SPI} \cite{SPI_Sensitivity}, and for an observation time of $10^5 \;\mathrm{s}$ for \emph{IBIS-ISGRI} and \emph{IBIS-PICsIT} \citep{IBIS_Sensitivity}. \emph{COMPTEL} sensitivity is given for the observation time accumulated during the $\sim 9\;\mathrm{years}$ duration of the \emph{CGRO} mission \cite{COMPTEL_Sensitivity}. The \emph{Fermi-LAT} sensitivity is given after 10 years of observation in survey mode, for a source at high Galactic latitudes\textsuperscript{2}. Concerning imaging atmospheric Cherenkov telescopes, the sensitivity is conventionally computed for an observation time $\mathrm{T_{obs}} = 50 \;\mathrm{hours}$ \cite{CTA_WhiteBook}. \emph{HAWC} and \emph{HiSCORE} sensitivities are shown after $5$ years of observations (see respectively \cite{HAWC_Sensitivity} and \cite{HiSCORE_Instrument}), while \emph{LHAASO} sensitivity after $1$ year of observations \cite{LHAASO_Instrument}. The curves for X-ray and soft gamma-ray instruments (from \emph{Chandra} up to \emph{COMPTEL}) are usually given in literature for a significance level of $3\,\sigma$; the others are given for a significance level of $5\,\sigma$.}
\small\textsuperscript{1} \url{http://integral.esac.esa.int/integ_payload_jemx.html}.\\
\small\textsuperscript{2} \url{https://www.slac.stanford.edu/exp/glast/groups/canda/lat_Performance.htm}.
\end{figure}

The measurement of X-ray and gamma-ray emission from astrophysical sources shed light on the most energetic processes in the Universe. X-ray and MeV gamma-ray astronomy can only be performed by space satellites (or balloons), due to complete photon absorption by the atmosphere at these energies. \emph{Wolter-type} telescopes \cite{WolterTelescope} and \emph{Coded-mask} telescopes \cite{CodedMask} are the most outstanding techniques to explore the X-ray and hard X-ray domain (up to hundreds of keV). The \emph{Chandra} X-ray observatory \cite{Chandra_Instrument}, \emph{NuSTAR} \cite{NuStar_Instrument}, \emph{XMM-Newton} \cite{XMM_Instrument} and \emph{INTEGRAL} \cite{INTEGRAL_Instrument} are some of the currently operating telescopes employing these technologies. The study of gamma rays in the energy range from $\sim50\,\mathrm{MeV}$ up to $\sim1\,\mathrm{TeV}$ is accomplished by pair conversion telescopes, like the \emph{Fermi-LAT} \cite{Atwood_FERMI} and the \emph{AGILE} \cite{Tavani_AGILE} satellites, continuously scanning the sky with a large field-of-view. At even higher energies the flux of astrophysical sources is generally too low to allow a significant statistic in the expected lifetime of a space telescope, given their limited effective area. Therefore observations of very-high-energy gamma rays are based on ground-based observatories, such as \emph{Imaging Atmospheric Cherenkov Telescopes} and \emph{Extensive Air Shower} observatories: \emph{MAGIC} \cite{MAGIC_Instrument}, \emph{H.E.S.S.} \cite{HESS_Instrument}, \emph{VERITAS} \cite{VERITAS_Instrument}, and the \emph{HAWC} observatory \cite{HAWC_Instrument} represents the current state-of-the-art of very-high-energy gamma-ray astronomy. Further development are expected in the near future, thanks to the construction and completion of observatories like \emph{LHAASO} \cite{LHAASO_Instrument}\footnote{The whole \emph{LHAASO} detector array has be completed in June 2021.}, \emph{CTA} \cite{CTA_WhiteBook} and \emph{HiSCORE} \cite{HiSCORE_Instrument}. Further improvements are expected in the X-ray and high-energy gamma-ray range, thanks to \emph{Athena} \cite{Athena_Intrument} and \emph{HERD} \cite{HERD_Instrument}.\\
This leaves the soft gamma-ray sky, from $\sim 200\,\mathrm{keV}$ to $\sim 50\,\mathrm{MeV}$, as one of the least explored regions of the electromagnetic spectrum. This gap in observations is often referred to as the ``MeV gap'' in the literature (Figure~\ref{fig:MeVGap}, highlighted by the grey region.). \emph{COMPTEL} \cite{COMPTEL_Instrument}, onboard \emph{CGRO}, opened the MeV gamma-ray band as a new window to astronomy, performing the first all-sky survey in the energy range from $0.75$ to $30\;\mathrm{MeV}$. However the performance of the telescope was quite modest when compared to the improvements achieved today in the other energy bands. More than $20$ years after the de-orbit of \emph{CGRO}, no dedicated successor mission is yet operating. The MeV energy range is currently covered only by the \emph{INTEGRAL} observatory, taking data since $2002$, but its sensitivity and field-of-view at $1\;\mathrm{MeV}$ is worse than that of \emph{COMPTEL} (see Figure~\ref{fig:MeVGap}). The scientific case for a deeper exploration of this energy range is strong and includes a broad selection of topics, from the search of electromagnetic counterparts of gravitational wave and neutrino events, to the study of emission mechanisms in blazars and the nucleosynthesis of heavy elements in our Galaxy. The reader interested in a comprehensive overview of the science case is referred to \cite{DeAngelis_eAstrogam_2018} and to selected sections of the \textit{Astro2020: Science White Papers} \footnote{\url{https://baas.aas.org/astro2020-science}}.

\noindent Many missions and concepts have been proposed in order to fill the MeV sensitivity gap, exploring different configurations and imaging techniques. \emph{AMEGO} \cite{AMEGO_instrument} is a probe-class mission concept, operating both as a Compton and pair telescope in order to achieve unprecedented sensitivity between $200\;\mathrm{keV}$ and $5\;\mathrm{GeV}$. \emph{GECCO} \cite{GECCO_instrument} is a novel concept to be proposed as a future \emph{NASA} Explorer mission, combining a Compton Cadmium Zinc Telluride telescope and a deployable coded aperture mask. \emph{ETCC} \cite{ETCC_instrument} is a Compton telescope concept based on a gaseous electron tracker and position-sensitive scintillation cameras. On a smaller scale, \emph{COSI} \cite{COSI_Instrument} is a Compton telescope employing germanium detectors, and was recently selected as a small explorer mission\footnote{\url{https://www.nasa.gov/press-release/nasa-selects-gamma-ray-telescope-to-chart-milky-way-evolution}}.\\
We propose here an alternative concept for a Compton telescope based on the CubeSat standard, named \emph{MeVCube}, with the advantages of low cost and relatively short development time. The proposed concept can provide a technology demonstrator, increasing the readiness level of hardware to be potentially implemented in future missions, while still providing its own scientific impact. A first evaluation of this concept has been presented in \cite{MeVCube_first}. However, there  sensitivities were estimated based on projections for the performance achievable with CdZnTe detectors. Here, we present a full set of performance metrics for the proposed telescope, such as effective area, field-of-view, energy resolution, angular resolution and sensitivity, obtained by simulations tuned to the measured properties of a prototype CdZnTe detector in a laboratory setup \cite{Lucchetta_CZT}.\\
\emph{MeVCube} will detect gamma rays in the energy range from $\sim200\;\mathrm{keV}$ to $\sim4\;\mathrm{MeV}$, and can therefore provide interesting insights into the ``MeV gap''. Spectral lines from the radioisotope decay due to nucleosynthesis in astrophysical sources typically have energies below $2\;\mathrm{MeV}$ \cite{Diehl_Spectroscopy}. The prompt emission from gamma ray bursts peaks in the energy range around hundreds of keV \cite{GRB_General}. Blazars typically exhibit a characteristic double-peaked spectral energy distribution (SED), with the high-energy peak of the most powerful blazars peaking in the keV-MeV range \cite{AGN_General}. \emph{MeVCube} can provide key information in the modelling of these sources, sampling their SED below or at the peak, which will be in particular valuable for multimessenger sources (see, e.g., ~\cite{Rodrigues:2018tku}). In addition, measurements of gamma-ray polarization would offer an important new technique for understanding the emission mechanisms in many of these objects. In a Compton telescope like \emph{MeVCube}, \emph{COSI} or \emph{AMEGO} polarization information can be retrieved from the distribution of the azimuthal scatter angle - i.e. the angle of the scattered photon with respect to the polarization plane of the incident radiation. Since gamma-ray bursts, and transient events in general, are very bright, even a small instrument with wide field-of-view such as \emph{MeVCube} can be effectively used for their detection.

\section{Design of a Compton Telescope based on the CubeSat standard}
\begin{figure}[htbp!]
\centering
\includegraphics[width=0.70\textwidth]{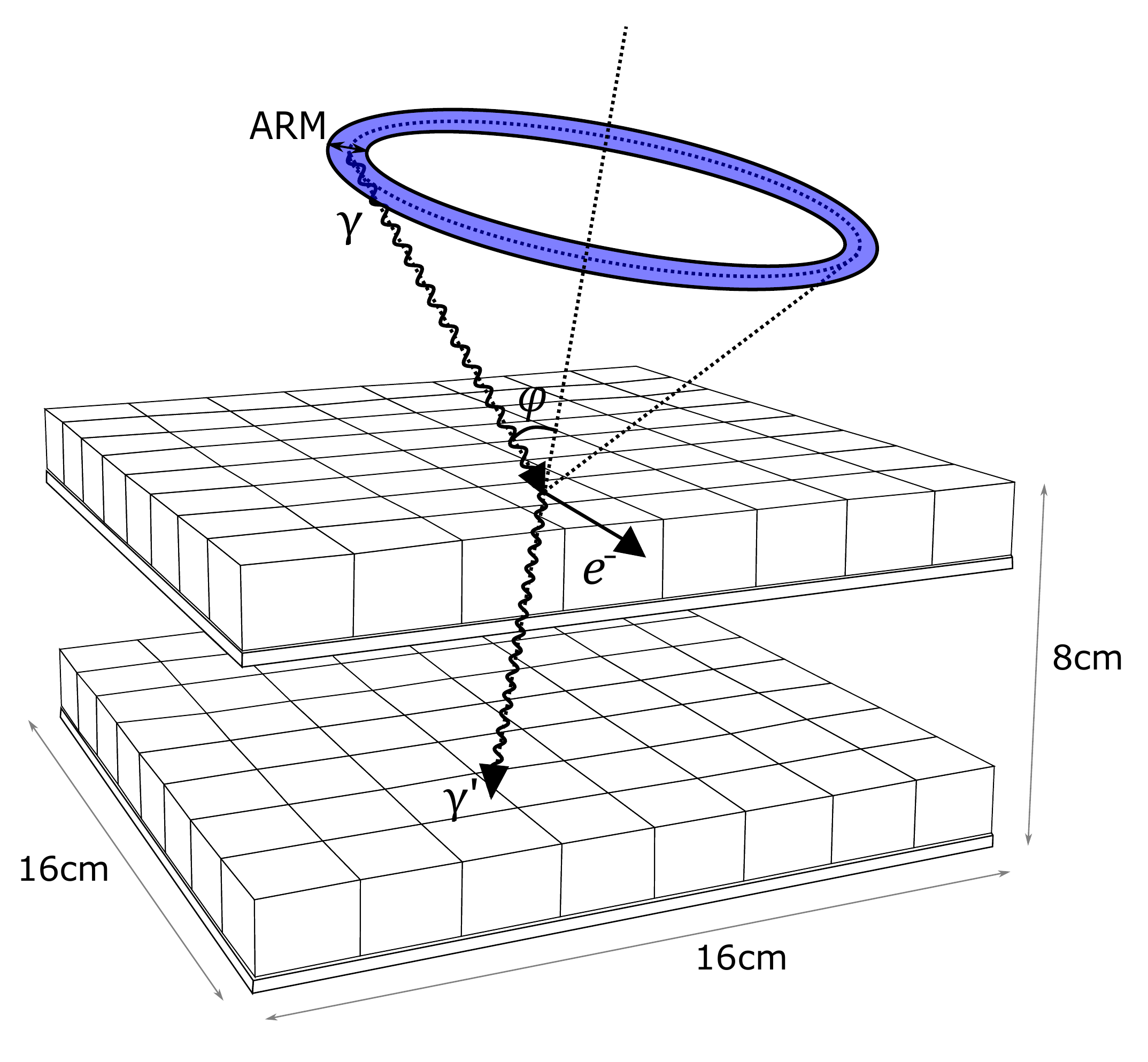}
\caption{\label{fig:MeVCubeGeo} Schematic mass model of the \textit{MeVCube} Compton telescope. The instrument implements $128$ CdZnTe detectors on two layers and an anti-coincidence detector (not shown in the figure). In a Compton camera, the direction of the incoming gamma ray is constrained to an annulus in the sky. The width of the annulus, quantified by the \emph{angular resolution measure} or ARM, defines the angular resolution of the Compton telescope.}
\end{figure}

A Compton telescope measures MeV gamma rays via Compton scattering in high Z detectors. Such a telescope provides excellent detection efficiency, wide field-of-view of the order of steradians and suitable angular resolution ($\mathcal{O}(1^{\circ})$). The payload proposed here is expected to match the constraints of a CubeSat mission, a class of nano-satellites with standardized size and form factor. A single CubeSat unit ($1\,\mathrm{U}$) has a volume of $10\times 10\times 11.35\;\mathrm{cm^3}$ and a maximum weight of $1.33\;\mathrm{kg}$. More than one unit can be combined together and the current CubeSat Design Specification defines the envelopes for $1\,\mathrm{U}$, $1.5\,\mathrm{U}$, $2\,\mathrm{U}$, $3\,\mathrm{U}$ and $6\,\mathrm{U}$ form factors \cite{CubeSatSpecification_2015}, with possible extensions to $12\,\mathrm{U}$ and $16\,\mathrm{U}$. \emph{MeVCube}'s scientific payload is expected to fit in an estimated volume of $20 \times 20 \times 10\;\mathrm{cm}^3$ (i.e. $\sim 4\,\mathrm{U}$). In this configuration, about $\sim 2\,\mathrm{U}$ would be required in the final design of the satellite for batteries, on-board electronics, reaction wheels, attitude control system etc., leading to an estimated form factor of $6\,\mathrm{U}$ in total. The \emph{MeVCube} design also fulfills the power of CubeSats, which is limited to just a few Watts.\\
The ``core'' of the instrument consists of $128$ Cadmium Zinc Telluride (CdZnTe or CZT) detectors, arranged in a stack of two layers of $64$ detector each, with a separation between the layers of $6\;\mathrm{cm}$ center to center (see Figure~\ref{fig:MeVCubeGeo})\footnote{The value of $6\;\mathrm{cm}$ has been chosen in order to match the size constraints given by the CubeSat standard and considering that additional space is needed for an anti-coincidence detector and the support structure.}. Soft gamma rays in the energy range from hundreds of keV up to few MeV predominately interact through Compton scattering in the first layer of the telescope (also referred as the scattering layer), while the scattered photon is in turn absorbed in the second layer (the absorption zone). According to the principle of operation of a Compton camera \cite{VonBallmoos_ComptonImaging,Boggs_ComptonReconstruction}, by measuring the energies of the recoil electron and of the scattered photon produced in the Compton process, together with their interaction positions, the direction of the incoming gamma ray can be constrained to an annulus in the sky:
\begin{equation}
\vec{e}\,'_{\gamma} \cdot \vec{e}_{\gamma} = \cos \varphi = 1 - \frac{m_e c^2}{E'_{\gamma}} + \frac{m_e c^2}{E'_{\gamma} + E_{e}} \; .
\label{eq:EventCycle}
\end{equation}
In Equation \eqref{eq:EventCycle}, $\vec{e}_{\gamma}$ and  $\vec{e}\,'_{\gamma}$ denote the direction of the incoming and scattered photon, respectively. $\varphi$ is the Compton scatter angle, and $E'_{\gamma}$ and $E_{e}$ are the energies of the scattered photon and the recoil electron. It follows that both, good energy and spatial resolutions, are needed in a Compton telescope in order to precisely reconstruct the origin of the incoming gamma rays. Cadmium Zinc Telluride (CdZnTe or CZT) semiconductor detectors \cite{Schlesinger_CZTreview} are ideally matched for applications on a Compton telescope. First of all, the high atomic number and density ensure a suitable stopping power and detection efficiency in the energy range of interest. Moreover, when combined with a fine segmented anode, they can also achieve great imaging and spectral performance at room temperature.

\noindent In our design, each CdZnTe detector has a volume of $2.0\times 2.0 \times 1.5\;\mathrm{cm}^3$, employing a $8 \times 8$ pixel anode structure and a planar cathode. The pixel pitch is $2.45\;\mathrm{mm}$ and the pixel size $2.25\;\mathrm{mm}$. For each event interaction, the triggered pixels provide information on the deposited energy and location on the anode plane (x-y plane), while the interaction depth (location on the z axis) is reconstructed from the ratio between the cathode signal and pixel signal. Pixels from each detector are read-out by the $64$-channel ASIC\footnote{Application Specific Integrated Circuit.} VATA450.3 developed by \emph{Ideas}, on the flip-side of the board hosting the detector itself. The read-out of cathode signals, based on VATA450.3 as well, is provided from a second board at the side of the payload, together with the power supply system. VATA450.3 has been selected based on its performance in terms of dynamic range, noise and linearity. Its low power consumption of only $0.25\;\mathrm{mW/channel}$ allows to operate the detectors under the power constrains present in CubeSats. Moreover, the ASIC has already been selected for space operation in the past \cite{Tajima_VATA450}. Performance of the read-out electronics and a custom designed detector have extensively been measured at DESY; the main results are presented in \cite{Lucchetta_CZT}. The main specifications and requirements for the \emph{MeVCube} Compton camera are summarized in Table~\ref{tab:MeVCUbeSpecifications}.\\
The Compton camera is enclosed on the top and the sides by an anti-coincidence detector (ACD, not shown in the schematic mass model of Figure~\ref{fig:MeVCubeGeo}), used to veto the in-orbit cosmic-ray background. The technology is similar to the one implemented in other previous missions, like \emph{Fermi-LAT} and \emph{Agile}.\\
The modular design of the CubeSat allows for a straightforward scaling into different payloads and form factors. For example a configuration that consists of $32$ CdZnTe detectors ($16$ per layer) could fit in a $2\,\mathrm{U}$ CubeSat model. Instead, a geometry employing $4$ layers of $64$ CdZnTe detectors each, with a separation of $5.5\;\mathrm{cm}$ center to center, matches a $12\,\mathrm{U}$ configuration \footnote{With the size of the scientific payload, also the size of the satellite bus changes, due to the different requirements for batteries, reactions wheels etc. We estimate that, in a $2\,\mathrm{U}$ configuration, $1\,\mathrm{U}$ can be dedicated to the scientific payload, while another unit will be dedicated to the satellite bus. Similarly in the $12\,\mathrm{U}$ configuration about $8\,\mathrm{U}$ are dedicated to the scientific payload and $4\,\mathrm{U}$ to the CubeSat bus.}.

\begin{table}[ht!]
\centering
\begin{tabular}{|l|l|}
\hline
\textbf{Parameter} & \textbf{Design value} \\
\hline
CubeSat model &  $4\,\mathrm{U}$ scientific payload, \\
 & $6\,\mathrm{U}$ complete satellite \\
\hline
Orbit & Low Earth Orbit (LEO), \\
 & $\sim 550\;\mathrm{km}$ altitude, $\leq 5^{\circ}$ inclination \\ 
\hline
Number of CdZnTe detectors & $128$ \\ 
\hline
CdZnTe detector size & $2.0\;\mathrm{cm} \times 2.0\;\mathrm{cm} \times 1.5\;\mathrm{cm}$ \\
\hline
Pixel pitch & $2.45\;\mathrm{mm}$ \\
\hline
Pixel size & $2.25\;\mathrm{mm} \times 2.25\;\mathrm{mm}$ \\ 
\hline
Depth resolution (FWHM) & $\sim 1.8\;\mathrm{mm}$ \\
\hline
Energy resolution (FWHM) & $\sim 6.5\%$ at $200\;\mathrm{keV}$, \\
 & $\sim 2.8\%$ at $662\;\mathrm{keV}$, \\
 & $\lesssim 2.0\%$ at $>1\;\mathrm{MeV}$ \\
\hline
Read-out electronics & VATA450.3 \\
\hline
Total power consumption & $<5\;\mathrm{W}$ \\
\hline
\end{tabular}
\caption{\label{tab:MeVCUbeSpecifications} Main specifications and requirements for the \textit{MeVCube} Compton telescope. Performance like the energy resolution and the spatial resolution are based on the measurements performed in \cite{Lucchetta_CZT}.}
\end{table}

\section{Evaluation of MeVCube performance through simulations}
The \emph{MeVCube} performance has been characterized in terms of the effective area, the field-of-view, the energy resolution, and the angular resolution. Detailed simulations have been performed with the \emph{Geant4} based toolkit \emph{MEGAlib} (\emph{Medium Energy Gamma-ray Astronomy Library}, \cite{Zoglauer_MEGAlib}), in order to evaluate the response of the instrument under point-like and monochromatic photon sources at different incidence angles, in the range from $\sim200\;\mathrm{keV}$ to $\sim4\;\mathrm{MeV}$. Originally developed inside the \emph{MEGA} project in the early 2000's \cite{MEGA_project}, the software has been maintained and further optimized through its use for other gamma-ray telescopes like \emph{eASTROGAM} \cite{deAngelis_eASTROGAM}, \emph{AMEGO} \cite{AMEGO_instrument} and \emph{COSI} \cite{COSI_Instrument}.

\subsection{Energy resolution}
\begin{figure}[htbp]
\centering
\includegraphics[width=0.70\textwidth]{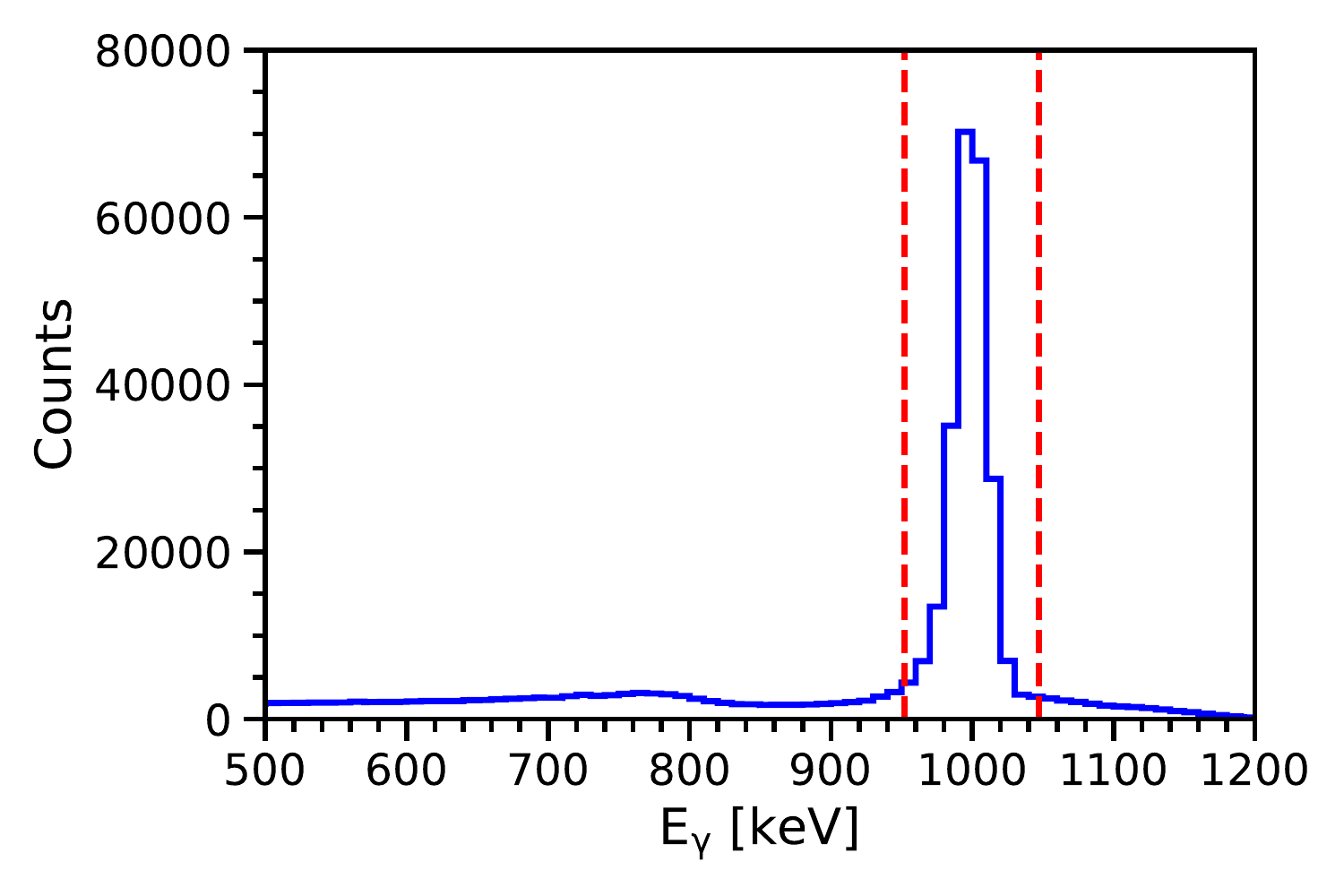}
\caption{\label{fig:MeVCube_EnergyDistr} Energy spectrum reconstructed from a simulation of a $1\;\mathrm{MeV}$ monochromatic photon source. The vertical dashed lines define the photo-peak in the region corresponding to $\pm2 \cdot\mathrm{FWHM}$ around the centroid.}
\end{figure}

The energy resolution of the telescope is given by the FWHM\footnote{Full width at half maximum.} of the reconstructed photo-peaks. As highlighted by \eqref{eq:EventCycle} a good spectral response is a fundamental requirement for a Compton telescope and is directly related to its angular resolution. The energy of the incident simulated photons is reconstructed from the sum of the energies of the recoil electrons and scattered photons, deposited inside the detectors. As it can be observed from Figure~\ref{fig:MeVCube_EnergyDistr}, the reconstructed spectrum is characterized by the photo-peak, here at $1\;\mathrm{MeV}$, followed by a long tail toward lower energy deposits, due to incomplete particle absorption and leakage. Obviously, the fraction of non-completely absorbed events increases with energy, and limits the \emph{MeVCube} energy range of operation.

\begin{figure}[htbp]
\centering
\includegraphics[width=0.70\textwidth]{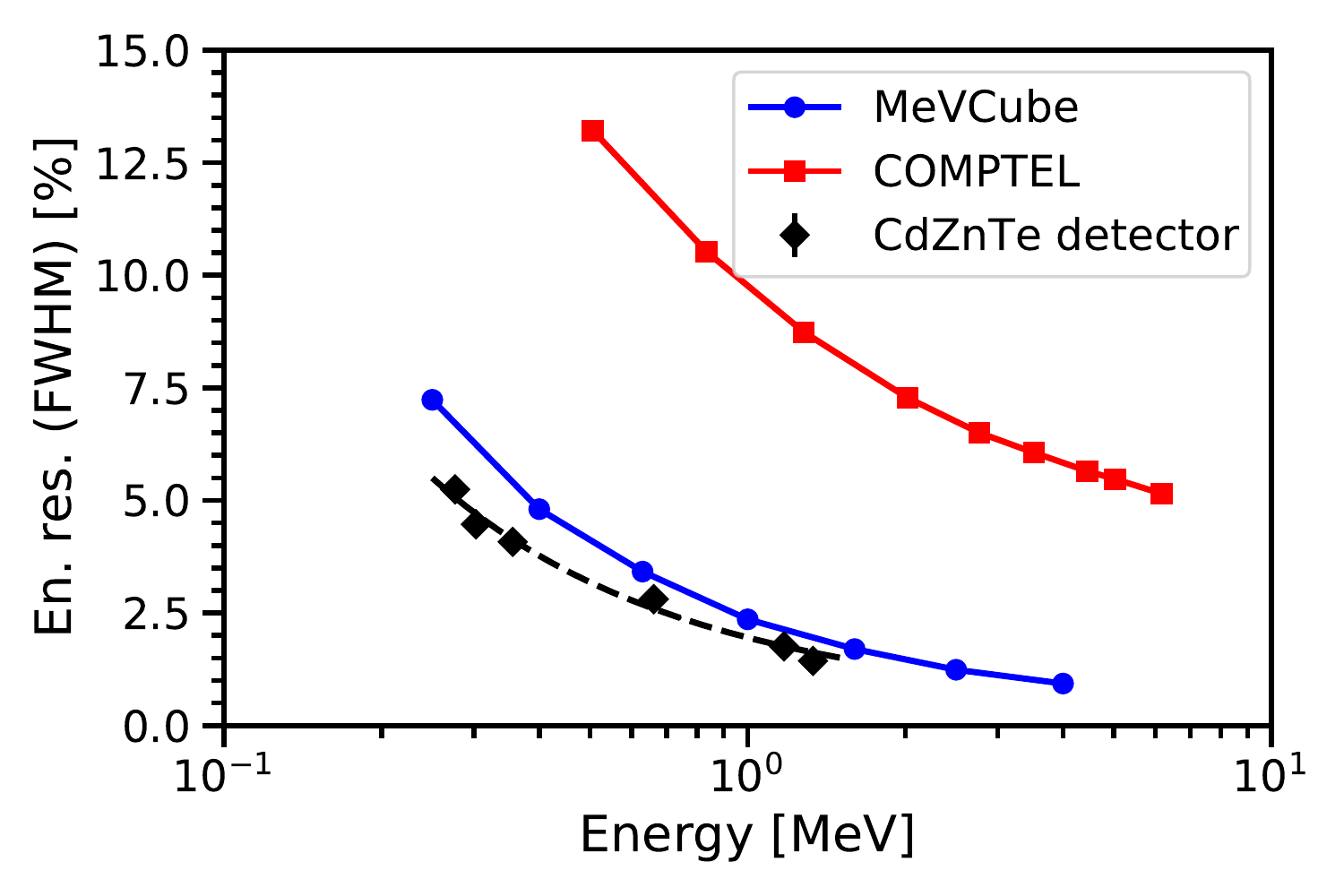}
\caption{\label{fig:MeVCube_EnergyRes} \emph{MeVCube} energy resolution compared to the one achieved by \emph{COMPTEL}. \emph{COMPTEL}'s data are taken from \cite{COMPTEL_Instrument}. The energy resolution of the single CdZnTe detector has been measured in \cite{Lucchetta_CZT}.}
\end{figure}

\noindent Figure~\ref{fig:MeVCube_EnergyRes} highlights the \emph{MeVCube} spectral performance as a function of the energy: an energy resolution of $\sim 2.5\%$ FWHM/E is achieved at $1\;\mathrm{MeV}$. In the Compton event reconstruction multiple particle interactions are recorded (at least two) in the detectors, and the errors add up for each interaction. Therefore the energy resolution of the telescope is slightly higher than the one measured for the single CdZnTe detector. As a direct consequence of the employment of semiconductor detectors over scintillation detectors, the spectral performance of \emph{MeVCube} is greatly improved compared to \emph{COMPTEL}.

\subsection{Effective area}
\begin{figure}[htbp]
\centering
\includegraphics[width=0.70\textwidth]{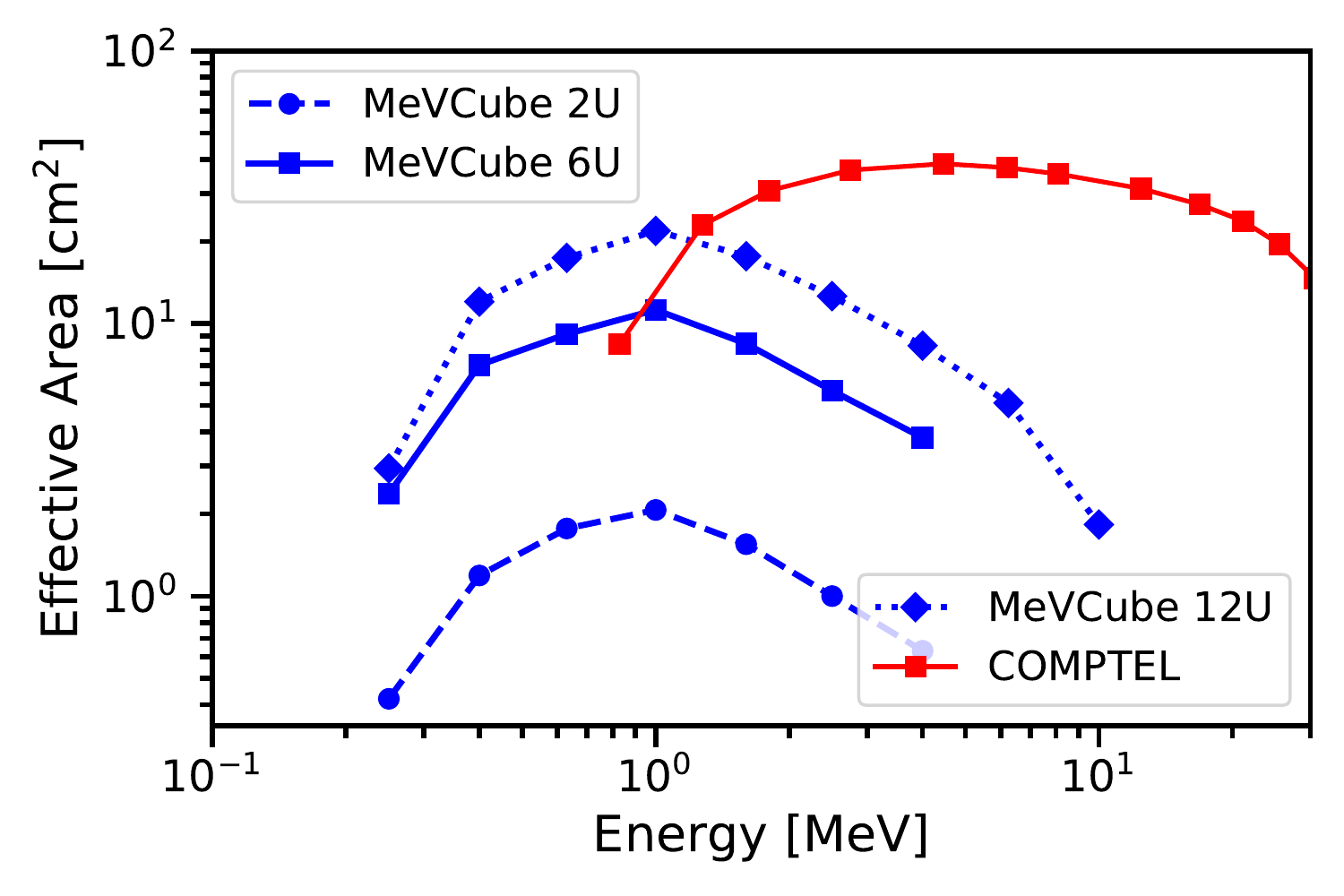}
\caption{\label{fig:MeVCube_Aeff} \textit{MeVCube} effective area as a function of the energy, compared to \emph{COMPTEL}. \emph{COMPTEL}'s data are taken from \cite{Kappadath_COMPTEL}.}
\end{figure}

The effective area of a telescope provides a measure of the instrument's detection efficiency, as a function of the photon energy and direction:
\begin{equation}
A_{eff} (\theta , \phi , E_\gamma) = A_{geo} (\theta , \phi) \cdot \varepsilon (\theta , \phi , E_\gamma)\; .
\label{eq:EffectiveArea}
\end{equation}
In this analysis the effective area is computed as the number of correctly reconstructed events, passing the selection cuts, divided by the number of simulated events and scaled by the simulation area surrounding the telescope. The effective area as a function of the energy, computed for vertically incident photons, is provided in Figure~\ref{fig:MeVCube_Aeff}. The impact of the different \emph{MeVCube} payloads considered is also addressed in the plot. For comparison \emph{COMPTEL}'s effective area, after quality cuts and zenith angles, was around $15\;\mathrm{cm^2}$ at $1\;\mathrm{MeV}$ and almost $40\;\mathrm{cm^2}$ at $5\;\mathrm{MeV}$.

\begin{figure}[htbp]
\centering
\includegraphics[width=0.70\textwidth]{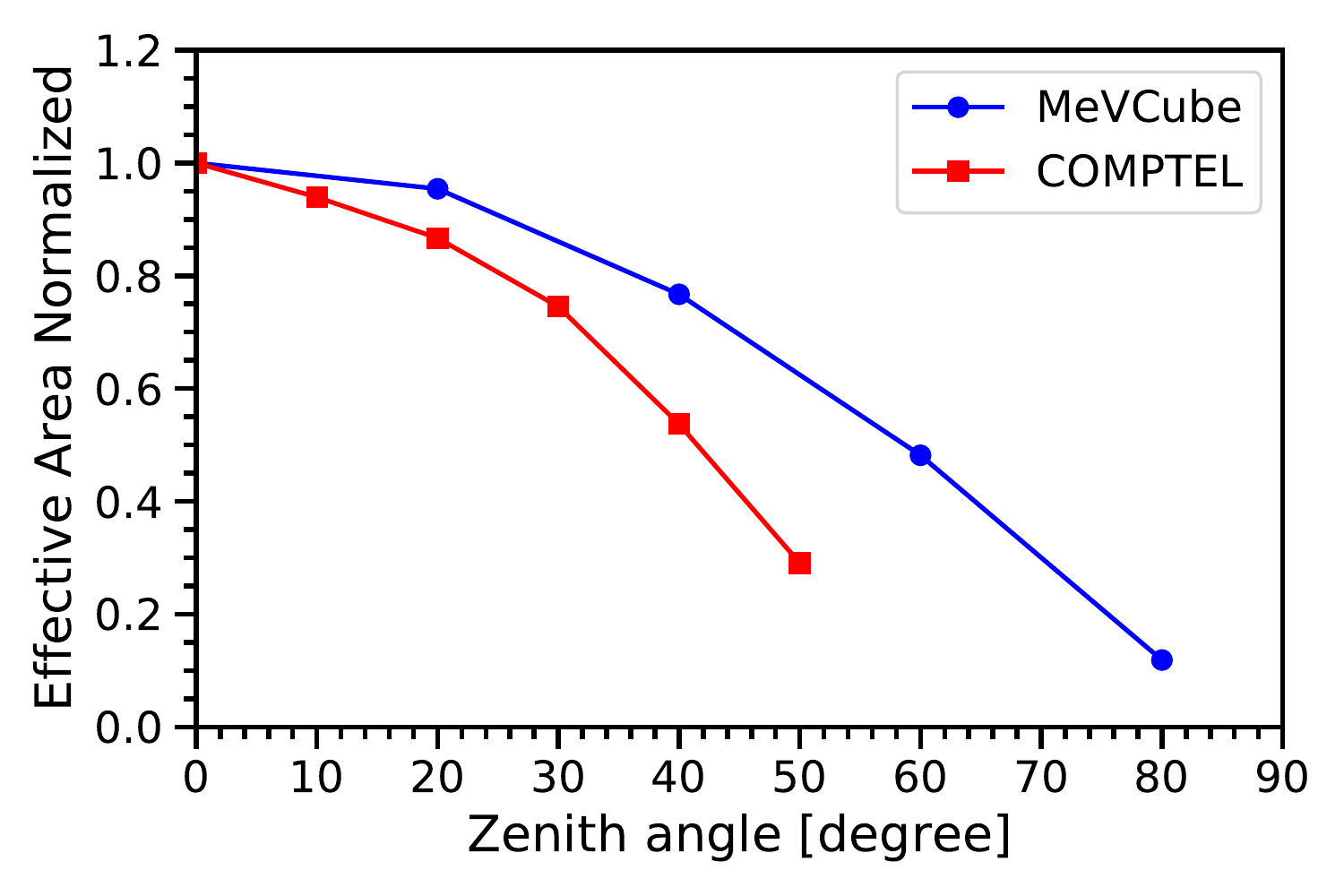}
\caption{\label{fig:MeVCubeFov} Comparison of the \emph{MeVCube} and \emph{COMPTEL} normalized effective areas as a function of the incident incident angle, at $1\;\mathrm{MeV}$. \emph{COMPTEL}'s data are taken from \cite{Kappadath_COMPTEL}.}
\end{figure}

\noindent In Figure~\ref{fig:MeVCubeFov} the effective area of \emph{MeVCube} and \emph{COMPTEL} are displayed as a function of the photon polar angle for an energy of $1\;\mathrm{MeV}$. The values have been normalized to the effective area for vertically incident photons, for a direct comparison of the two instruments. \emph{MeVCube}'s field-of-view is wider than the \emph{COMPTEL} one (almost $\sim 2\;\mathrm{sr}$ compared to $\sim 1\;\mathrm{sr}$). The acceptance of \emph{MeVCube}, i.e. the effective area integrated over the solid angle, at $1\;\mathrm{MeV}$ is $21\;\mathrm{cm^2\,sr}$ in the $6\,\mathrm{U}$ and $42\;\mathrm{cm^2\,sr}$ in the $12\,\mathrm{U}$; \emph{COMPTEL}'s acceptance was $16\;\mathrm{cm^2\,sr}$ at $1\;\mathrm{MeV}$.

\subsection{Angular resolution}
\begin{figure}[htbp]
\centering
\includegraphics[width=0.70\textwidth]{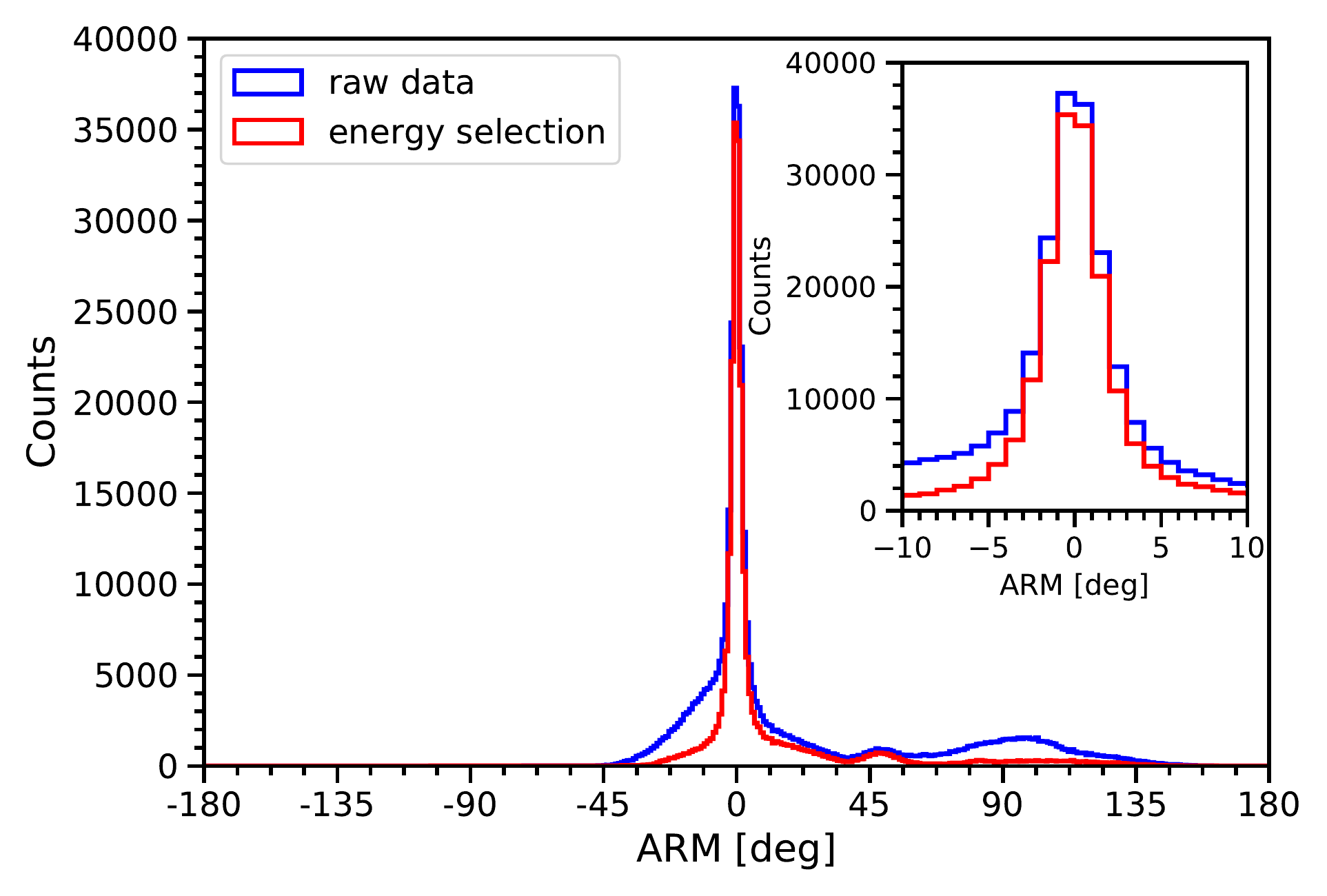}
\caption{\label{fig:MeVCube_ARMdistr} Distribution of the ARM from a simulation of $1\;\mathrm{MeV}$ photons, before and after selection of events in the full energy photo-peak (blue and red histograms respectively). The inset of the figure is provided for better visibility of the peak.}
\end{figure}

The angular resolution of a Compton telescope is canonically quantified by the \emph{angular resolution measure} or ARM. The ARM (see also Figure~\ref{fig:MeVCubeGeo}) is defined as the smallest angular distance between the nominal source location and each event annulus:
\begin{equation}
ARM = \arccos(\vec{e}\,'_{\gamma} \cdot \vec{e}_{\gamma}) - \varphi \, .
\label{eq:ARM}
\end{equation}
Therefore the ARM distribution represents the point spread function of a Compton telescope, providing a measure of the uncertainty in the opening angle of the Compton cone for each reconstructed event.\\
The ARM distribution for simulated photons with an energy of $1\;\mathrm{MeV}$ is shown in Figure~\ref{fig:MeVCube_ARMdistr}. The distribution exhibits a narrow peak with tails due to incompletely absorbed scattered gamma rays and recoil electrons. In this context we point out that the selection of events in the photo-peak of Figure~\ref{fig:MeVCube_EnergyDistr}, highlighted by the dashed red lines, leads to a strong reduction of the tails, as evident from the red histogram in Figure~\ref{fig:MeVCube_ARMdistr} \footnote{We note that such a selection cannot be directly performed in a real life instrument.}. 
 
\begin{figure}[htbp]
\centering
\includegraphics[width=0.70\textwidth]{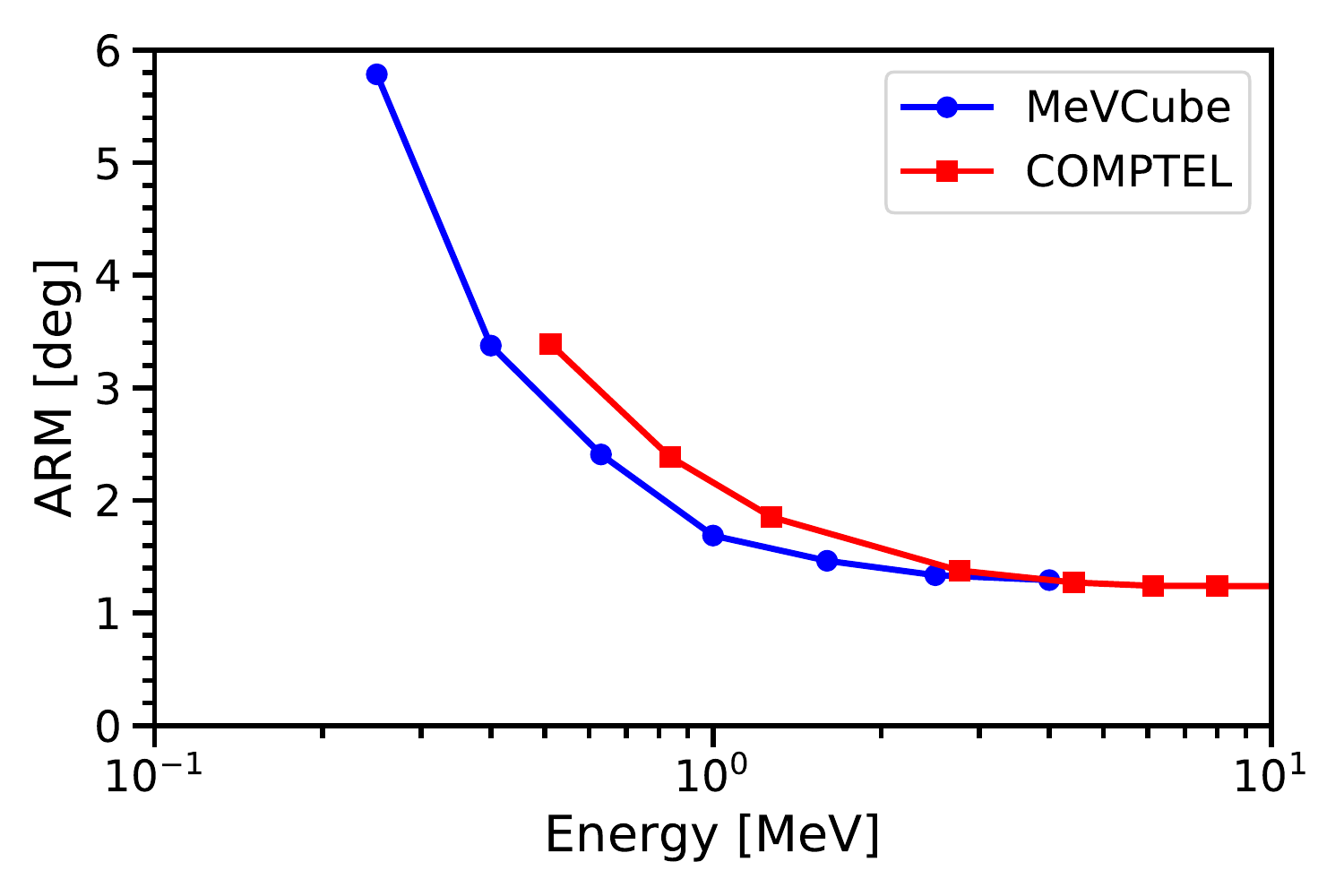}
\caption{\label{fig:MeVCube_AngularResolution} \emph{MeVCube} ARM as a function of the photons energy. \emph{COMPTEL} ARM \cite{COMPTEL_Instrument} is also reported, for comparison.}
\end{figure}

\noindent The angular resolution, computed as $1\,\sigma$ of the ARM distribution, is reported as a function of the photon energy in Figure~\ref{fig:MeVCube_AngularResolution}: an angular resolution of about $1.5^{\circ}$ is achieved at $1\;\mathrm{MeV}$. Despite the superior spatial and energy resolution of CdZnTe detectors, compared to scintillation detectors, \emph{MeVCube} angular resolution is only comparable to the one achieved by \emph{COMPTEL}. This is a consequence of the limited size of CubeSats: the separation between the two detector layers in \emph{MeVCube} is only $6\;\mathrm{cm}$, compared to $1.5\;\mathrm{m}$ in \emph{COMPTEL}.

\section{\textit{MeVCube} continuum sensitivity}
As high-level performance measure for the telescope, the continuum sensitivity is presented, quantifying the instrument's ability to detect faint sources. The sensitivity is evaluated based on the telescope performance for angular resolution, effective area and observation time, and the background rate. In this work, the sensitivity is computed for gamma-ray sources at high Galactic latitude and at high zenith angles with respect to the Earth. Before presenting \emph{MeVCube} sensitivity, we briefly summarize the instrumental and astrophysical background sources for a Compton telescope in a low-Earth orbit.

\subsection{Background sources for a Compton telescope on a low-Earth orbit}
The ideal orbit for \textit{MeVCube}, like other Compton telescopes, is a low-Earth orbit, with an altitude $\sim 550\;\mathrm{km}$ and nearly equatorial (inclination $\leq 5^{\circ}$). In \cite{Cumani_Background} a comprehensive model of the background sources for a gamma-ray telescope in a low-Earth orbit is provided. Different background components have been modelled in the energy range between $10\;\mathrm{keV}$ and $100\;\mathrm{GeV}$, based on measurements from instruments like \emph{Fermi-LAT}, \emph{INTEGRAL} and \emph{AMS}. Here, the main background components for \emph{MeVCube} are briefly outlined:
\begin{itemize}
\item \textbf{Extra-galactic gamma-ray background (EGB)}: a diffuse and isotropic photon background, due to unresolved sources \cite{Ajello_EGB}. A comprehensive study of the EGB in the MeV energy range will also be an important science topic for a future Compton telescope \cite{Ajello_MeVBackground}.
\item \textbf{Earth’s gamma-ray emission}: often also referred to as Earth's albedo in literature, is a strongly anisotropic gamma-ray flux, generated by the interaction between primary cosmic rays and the Earth's atmosphere. For a $550\;\mathrm{km}$ altitude low-Earth orbit, Earth's albedo emission peaks at a polar angle of $\sim113^{\circ}$ from zenith. Even if no significant emission is expected for angles $<90^{\circ}$, Earth's albedo background events still overlap with signal events in a Compton telescope, since their origin can only be reconstructed to a great circle on the sky.
\item \textbf{Charged-particle background}: primary and secondary cosmic rays constantly hit the detectors, contributing to the total background rate. In the energy range of interest for \emph{MeVCube} the major contribution comes from secondary positrons and electrons. However the majority of the charged-particle background can be effectively vetoed by the anti-coincidence detector and have not been considered for this study.
\item \textbf{Material activation}: the flux of cosmic rays and gamma rays, constantly hitting the spacecraft, activates the satellite materials, producing radioactive isotopes. Eventually, an equilibrium between the production of new radioactive nuclei and their decays to stable elements is reached. The energy spectrum of the instrumental background produced by the decay of radioactive isotopes, is described by a continuum emission plus several characteristic lines, with a strong overlap with the measured signals in a Compton telescope. Moreover the total on-orbit background rate depends strongly on the materials and payload of the spacecraft, and its orbit. An equatorial orbit, avoiding the passage through the South Atlantic Anomaly is preferable in this context.\\
In the case of \emph{COMPTEL}, the instrumental background was the major contribution for energies below $4.2\;\mathrm{MeV}$ \cite{COMPTEL_Sensitivity}. For a small satellite like \emph{MeVCube}, the instrumental background is expected to be lower than the one measured for large-scale missions, scaling approximately as mass$^{1/3}$. Material activation is neglected in the first order for the evaluation of \emph{MeVCube} sensitivity, but should be considered in future once all the details of the payload and orbit are finalized. 
\end{itemize}
As in the simulations performed in the previous section, we also require in the case of background a Compton scattering in the first layer and absorption of the scattered photon in the second layer. For such a trigger, soft EGB or Albedo photons which interact and are absorbed directly in the first layer are automatically rejected from our analysis.

\subsection{Continuum sensitivity evaluation}
For point-like gamma-ray sources the continuum sensitivity can be expressed by:
\begin{equation}
F_{z} = \frac{z^2 + z \sqrt{z^2 + 4 N_{\mathrm{bkg}}}}{2 T_{\mathrm{obs}} A_{\mathrm{eff}}} \, .
\label{eq:Sensitivity} 
\end{equation}
In \eqref{eq:Sensitivity}, $z$ is the statistical significance in unit of sigmas (here we consider a $3\,\sigma$ source detection), $T_{\mathrm{obs}}$ is the total observation time, $A_{\mathrm{eff}}$ is the effective area, and $N_{\mathrm{bkg}}$ is the number of background photons that lie within the angular resolution element defined by the telescope.\\
We note that in the energy range of interest the sensitivity is background dominated and therefore scales as the inverse of the square root of the observation time.

\begin{figure}[htbp]
	\centering
	\subfloat[\emph{MeVCube} sensitivity for an observation time of $10^5\;\mathrm{s}$, compared to the one achieved by the \emph{IBIS} and \emph{SPI} instruments on-board the \emph{INTEGRAL} observatory.]{
	\includegraphics[width=0.75\textwidth]{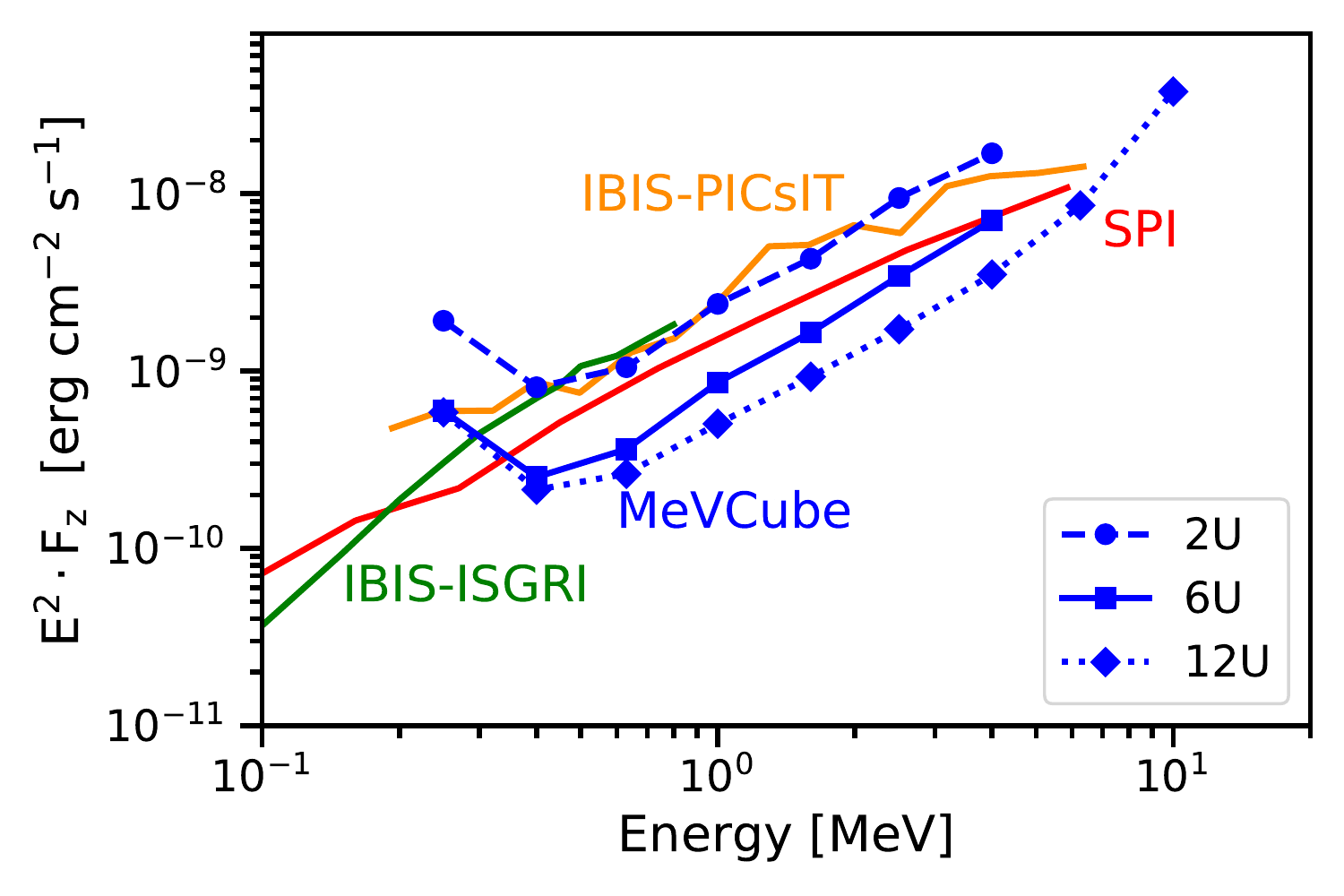}
	\label{fig:MeVCubeSens_comparisonINTEGRAL}}
	\quad
	\subfloat[\emph{MeVCube} sensitivity for an observation time of $2$ months, compared to the one achieved by \emph{COMPTEL}.]{
	\includegraphics[width=0.75\textwidth]{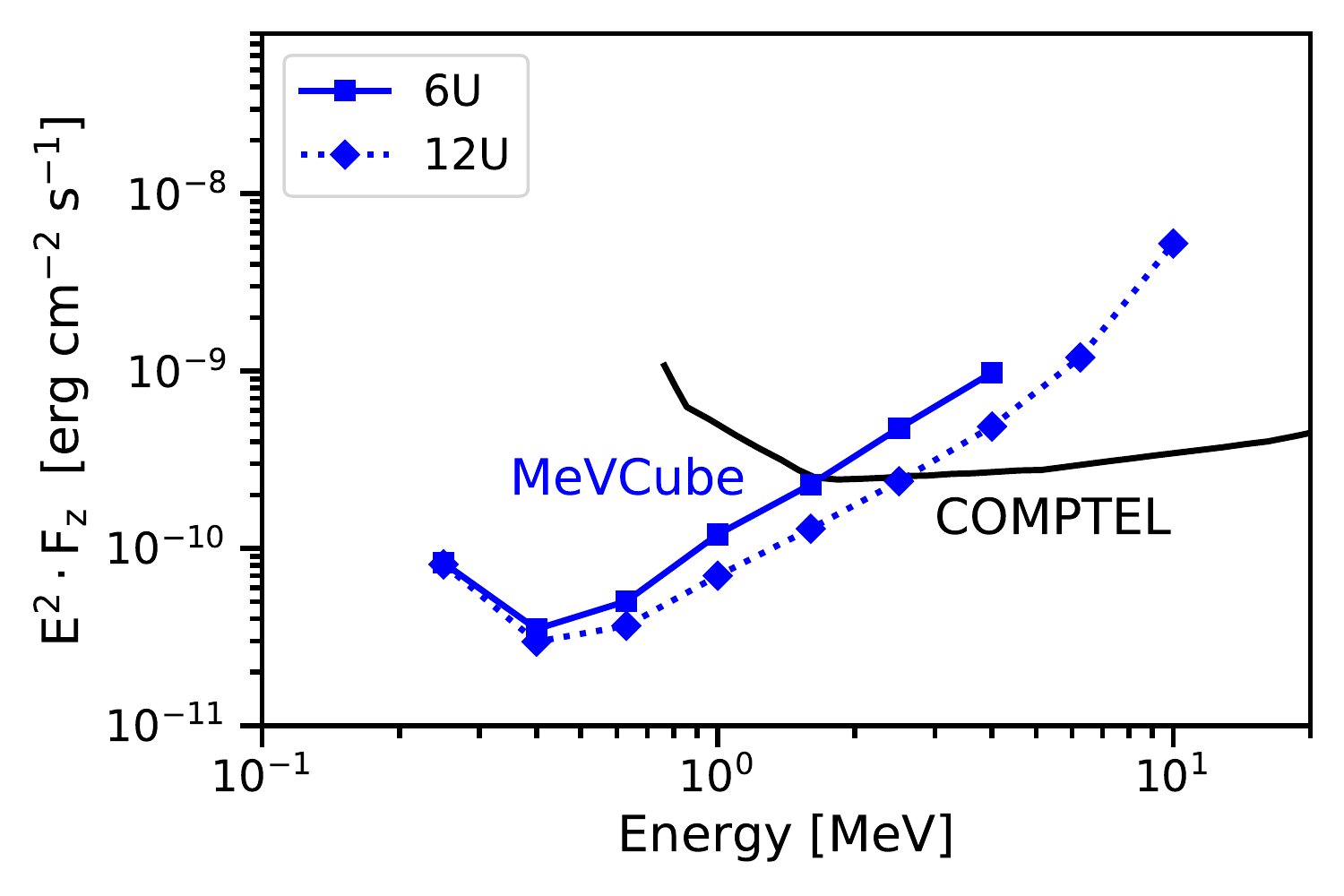}
	\label{fig:MeVCubeSens_comparisonCOMPTEL}}
	\caption{\label{fig:MeVCubeSens} \emph{MeVCube} continuum sensitivity compared to the one achieved by \emph{COMPTEL} and by the \emph{IBIS} and \emph{SPI} instruments on-board the \emph{INTEGRAL} observatory. \emph{COMPTEL} sensitivity is provided for the observation time accumulated during the $\sim 9\;\mathrm{years}$ duration of the \emph{CGRO} mission \cite{COMPTEL_Sensitivity}. Concerning the \emph{INTEGRAL} observatory, the sensitivity is computed for an observation time of $10^6\;\mathrm{s}$ for \emph{SPI} \cite{SPI_Sensitivity}, and for an observation time of $10^5\;\mathrm{s}$ for \emph{IBIS-ISGRI} and \emph{IBIS-PICsIT} \cite{IBIS_Sensitivity}.}
\end{figure}

\noindent We present \emph{MeVCube} sensitivity computed for the optimal case of gamma-ray sources at high latitude and zenith angle. Figure~\ref{fig:MeVCubeSens} compares the results of the simulation for different form factors, with the sensitivity achieved by \emph{COMPTEL} and \emph{INTEGRAL}. In the top, \emph{MeVCube} sensitivity has been computed for a continuous observation time of $10^5\;\mathrm{s}$ (corresponding to an effective observation time of $\sim1$ day), in comparison to the sensitivity achieved by the \emph{IBIS} and \emph{SPI} instruments on-board the \emph{INTEGRAL} observatory. In the $6\,\mathrm{U}$ configuration, \emph{MeVCube} reaches a sensitivity at $1\;\mathrm{MeV}$ of $8.6\cdot 10^{-10}\;\mathrm{erg\,cm^{-2}\,s^{-1}}$ or $8.1\cdot 10^{-4}\;\mathrm{ph\,cm^{-2}\,s^{-1}}$ for a $10^5\;\mathrm{s}$ observation time.\\
For a $6\,\mathrm{U}$ or a $12\,\mathrm{U}$ configuration, the expected mission lifetime could be a few years; given the field-of-view of the instrument and operation in survey mode, this would imply a $\sim 2$ months effective stare at the source. Therefore in the bottom of Figure~\ref{fig:MeVCubeSens}, the \emph{MeVCube} sensitivity has been computed for an observation time of $2$ months, in comparison to the one achieved by \emph{COMPTEL} during the $\sim 9\;\mathrm{years}$ duration of the \emph{CGRO} mission \cite{COMPTEL_Sensitivity}. During the operation of \emph{CGRO}, \emph{COMPTEL} performed about $340$ pointings of roughly $2$ weeks duration each. The sky coverage was not uniform on the entire sky, with effective observation time typically around $2-3$ months depending on the region; a sky exposure map of \emph{COMPTEL} can be found in \cite{COMPTEL_reloaded}.\\
\emph{MeVCube} can cover the energy range between $\sim 200\;\mathrm{keV}$ and $4\;\mathrm{MeV}$ with a sensitivity comparable to the last generation of large-scale MeV missions. The larger field-of-view and much lower cost are fundamental for observation of transients and gamma-ray bursts, where sky coverage is a key element.

\section{Conclusions and Outlook}
In this paper we presented the first performance evaluation of a novel Compton telescope concept based on the CubeSat standard, named \emph{MeVCube}. The impact of different sizes for the CubeSat payload was also addressed in our simulations. Despite the small effective area, limited to $\lesssim 10\;\mathrm{cm^2}$, \emph{MeVCube} can reach an energy resolution of $2.5\%$ FWHM/E and an angular resolution of $1.5^{\circ}$ at $1\;\mathrm{MeV}$. Given such a performance, \emph{MeVCube} can reach a sensitivity comparable to the one achieved by the last generation of large satellites like \emph{COMPTEL} and \emph{INTEGRAL}, in the energy range between $\sim 200\;\mathrm{keV}$ and $\sim 4\;\mathrm{MeV}$. Therefore, if launched into space, \emph{MeVCube} provides attractive scientific performance complementary to an instrument like \emph{COSI} while guaranteeing larger sky coverage. In particular the combination of wide field-of-view, reasonable angular resolution and comparatively low costs make \emph{MeVCube} a powerful instrument for observations of transients and searches of electromagnetic counterparts of gravitational wave events. Finally, such a Compton camera represents a technology demonstrator, increasing the readiness level of hardware to be implemented in future larger-scale missions.\\
Future work on refining the performance expectations for\emph{MeVCube} will include dedicated studies on the polarization measurement capabilities and the sensitivity to nuclear lines, which are both relevant in this energy range. Another interesting subject for further study is transient event detection outside of the nominal field-of-view. This seems possible with the \emph{MeVCube} design employing a similar concept as the one reported in \cite{NITRATES}, using events detected in only one of the detector planes. These events would greatly increase the field-of-view and effective area for transients at the expense of localization performance.\\

\acknowledgments
The authors would like to thank Ingo Bloch for the useful discussions and feedback about \emph{MeVCube} idea and concept. This publication is funded by the Deutsche Forschungsgemeinschaft (DFG, German Research Foundation) - 491245950.

% The bibliography will probably be heavily edited during typesetting.
% We'll parse it and, using the arxiv number or the journal data, will
% query inspire, trying to verify the data (this will probalby spot
% eventual typos) and retrive the document DOI and eventual errata.
% We however suggest to always provide author, title and journal data:
% in short all the informations that clearly identify a document.

\end{document}